\documentclass[aps,preprint]{revtex4-1}

\usepackage{amsmath,graphicx,bbm,mathrsfs,amssymb,pst-all,bm,color}

\begin{document}

\title{Quantum optimization algorithm based on multistep quantum computation}
\author{Hefeng Wang$^1$}
\email{wanghf@mail.xjtu.edu.cn}
\author{Hua Xiang$^2$}
\email{hxiang@whu.edu.cn}
\affiliation{$^{1}$ MOE Key Laboratory for Nonequilibrium Synthesis and Modulation of Condensed Matter, Xi'an Jiaotong University, Xi'an, 710049, China \\
School of Physics, Xi'an Jiaotong University and Shaanxi
Province Key Laboratory of Quantum Information and Quantum Optoelectronic
Devices, Xi'an, 710049, China}
\affiliation{$^{2}$School of Mathematics and
Statistics, Wuhan University and Hubei Key Laboratory of Computational Science, Wuhan, 430072, China}

\begin{abstract}
We present a quantum algorithm for finding the minimum of a function based on
multistep quantum computation and apply it for optimization problems with continuous
variables, in which the variables of the problem are discretized to form the
state space of the problem. Usually the cost for solving the problem
increases dramatically with the size of the problem. In this algorithm, the
dimension of the search space of the problem can be reduced exponentially
step by step. We construct a sequence of Hamiltonians such that the search
space of a Hamiltonian is nested in that of the previous one. By applying
a multistep quantum computation process, the optimal vector is finally
located in a small state space and can be determined efficiently. One of the
most difficult problems in optimization is that a trial vector is trapped
in a deep local minimum while the global minimum is missed, this problem
can be alleviated in our algorithm and the runtime is proportional to the
number of the steps of the algorithm, provided certain conditions are satisfied.
We have tested the algorithm for some continuous test functions.
\end{abstract}

\maketitle

\section{Introduction}

Optimization problem is one of the most important problems in science and
engineering. It includes a wide class of problems ranging from molecular
modeling, quantum mechanical calculations, machine learning, to
combinatorial optimization. These problems can be classified into different
categories, e.g., continuous or discrete optimization, constrained or
unconstrained optimization, convex or nonconvex optimization, differentiable
or nondifferentiable optimization, deterministic or stochastic optimization~%
\cite{weise, boyd, noce, flet, flou}, etc. There is no universal
optimization algorithm. Most classical optimization algorithms start with a
trial vector that is varied by using different techniques to find the
optimum of an objective function. The cost of the algorithms can become very
expensive due to the increase of the dimension of the state space of the
problem, which is known as \textquotedblleft the curse of
dimension\textquotedblright. Another problem that often happens for
optimization algorithms is that the trial vector is trapped in a deep local
minimum, while missing the global minimum of the objective function.

Optimization has also been studied in the framework of quantum computation.
Adiabatic quantum computing~(AQC) is designed for solving combinatorial
optimization problems~\cite{farhi}, in which starting with the ground state
of a simple initial Hamiltonian, the system is evolved adiabatically to a
final Hamiltonian whose ground state encodes the solution to the
optimization problem. Despite the theoretical guarantee of the adiabatic
theorem, the condition of adiabaticity in AQC is difficult to maintain in
practice, since the allowed rate of evolution is determined by the minimum
energy gap between the ground and the first excited states of the adiabatic
evolution Hamiltonian, which is not known \textit{a priori}. Quantum
annealing is a heuristic quantum optimization algorithm~\cite%
{fin,kad,bro,san,joh} that can be viewed as a relaxation of AQC, where the
conditions of adiabaticity are not met and the evolution time from an
initial Hamiltonian to the final Hamiltonian is determined heuristically.
Whether or not quantum annealing can provide quantum speed-up over classical
heuristic algorithms is still not clear. Variational quantum algorithms such
as quantum approximate optimization algorithm~(QAOA)~\cite{qaoa} are hybrid
quantum-classical algorithms designed for near-term noisy intermediate-scale
quantum computers~\cite{nisq} without performance guarantees. It is known
that in the infinite depth limit, the QAOA recovers adiabatic evolution and
would converge to the optimal solution. The gradient decent methods are used
for optimization problems with continuous variables. The methods find local
minima of a smooth function by moving along the direction of the steepest
descent. Quantum algorithm provides an efficient way in calculating
numerical gradients~\cite{jor}, and has been used in iterative algorithms
for polynomial optimization~\cite{reb}. Optimization algorithms based on
gradient decent require that the objective function to be smooth, and they
have the problem of being trapped in a local minimum and missing the global
minimum. Besides, as the dimensionality of the problem increases, the search
of the phase space becomes more and more complicated, and the complexity of
the algorithm increases. Another approach for continuous optimization is by
using Grover's search algorithm~\cite{grover}. Continuous optimization
problems can be discretized and mapped to a search problem, thereby solved
by using Grover's algorithm. The Grover adaptive search algorithms
iteratively apply Grover search to find the optimum value of an objective
function~\cite{durr,pro,bul,bari,kowa,liu,gill}, and can achieve quadratic
speedup over classical search algorithms. However, these brute force methods
are prohibitively expensive due to the large search space of the problems.

In a recent work~\cite{wyx}, we proposed an efficient quantum algorithm for
solving a search problem with nested structure through multistep quantum
computation. The problem can be decomposed and the search space of the
problem can be reduced in a polynomial rate. The runtime of the algorithm is
proportional to the number of steps of the algorithm. In this work, we
generalize this algorithm for optimization problems with continuous
variables.

The nested structured search problem~\cite{wyx} is a search problem that
contains $N$ items with one target item, and can be decomposed by using $m$ [%
$O(\log N)$] oracles to construct $m$ Hamiltonians, respectively, as
\begin{equation}
H_{P_{i}}=-\sum_{\eta _{i}\in \Pi _{i}}|\eta _{i}\rangle \langle \eta _{i}|,%
\text{ \ \ }i=1,\ldots ,m
\end{equation}%
and%
\begin{equation}
H_{P_{m}}=H_{m}=H_{P}=-|\eta \rangle \langle \eta |,
\end{equation}%
where the set $\Pi _{i}$ contains $N_{i}$ marked items in the $N$ items and $%
|\eta _{i}\rangle $ are the marked states associated with the marked items,
and $|\eta \rangle $ is the target state that defines the problem
Hamiltonian of the search problem. These sets are nested as $\Pi _{1}\supset
\cdots \supset \Pi _{m-1}\supset \Pi _{m}$ with sizes $N_{1}$, $\cdots $, $%
N_{m-1}$, $N_{m}=1$, respectively. The ratio $N_{i-1}/N_{i}$ are polynomial
large, and $N_{0}=N$. The goal is to find the the target state $|\eta
\rangle $ that is associated with the target item in the set $\Pi _{m}$.

Our algorithm solves the nested structured search problem by finding the
ground state of the problem Hamiltonian $H_{P}$ via a multistep quantum
computation process, which is realized through quantum resonant
transition~(QRT)~\cite{whf0, whf2}. In this algorithm, a probe qubit is
coupled to an $n$-qubit register $R$ that represents the problem. We
construct a sequence of intermediate Hamiltonians to form a Hamiltonian
evolution path to the problem Hamiltonian as
\begin{equation}
H_{i}=\frac{N_{i}}{N}H_{0}+\left( 1-\frac{N_{i}}{N}\right) H_{P_{i}},\text{
\ \ }i=0,1,\ldots ,m-1,
\end{equation}%
where $H_{0}=-|\psi _{0}\rangle \langle \psi _{0}|$ and $|\psi _{0}\rangle =%
\frac{1}{\sqrt{N}}\sum_{j=0}^{N-1}|j\rangle $. Then we start from the ground
state of the initial Hamiltonian, and evolve it through the ground states of
the intermediate Hamiltonians sequentially through QRT to reach the ground
state of the problem Hamiltonian. The ground state of an intermediate
Hamiltonian is protected in an entangled state of the probe qubit and the
register $R$, such that it can be used repeatedly without making copies.
Therefore the algorithm circumvents the restriction of the no-cloning
theorem~\cite{noclone1, noclone2} and realizes the multistep quantum
computation. The algorithm can be run efficiently provided that: ($i$) the
energy gap between the ground and the first excited states of each
Hamiltonian $H_{i}$ and, ($ii$) the overlaps between ground states of any
two adjacent Hamiltonians are not exponentially small. For the nested
structured search problem, the conditions of the algorithm are satisfied
since the ratio $N_{i-1}/N_{i}$ are polynomial large, therefore it can be
solved efficiently, and the conditions for efficiently running our algorithm
are not equivalent to those of the AQC algorithms~\cite{wyx}.

In this algorithm, by using the Hamiltonians $H_{P_{i}}$ sequentially in
each step, the dimension of the search space of the problem is reduced in a
polynomial rate, the solution state to the problem Hamiltonian is obtained
step by step. The idea of reducing the search space in a polynomial rate
step by step in our algorithm has a classical analogue as follows: suppose
there are $80$ balls, all of them have equal weights except one that is
lighter than the others. How to find the lighter ball? If we randomly pick
up a ball and compare its weight with the other balls, this will take about $%
40$ trials on average. If we have a balance, then how many times do we have
to use the balance to find the lighter ball? According to information
theory, the number of times the balance has to be used is $\log 80/\log
3\approx 4$. The procedure is as follows: we divide all the $80$ balls into $%
3$ groups, each group has $27$, $27$ and $26$ balls, respectively; then pick
up the two groups that both have $27$ balls, and use the balance to
determine if they have equal weights. If the answer is positive, pick the
group with $26$ balls and divide it into $3$ groups again: $9,9,8$;
otherwise, take the group that is lighter and divide it into three new
groups: $9,9,9$. This process can be repeated until the lighter ball is
found. In this example, we can see that the problem is divided into a series
of nested sub-problems and the size of the search space is reduced in a rate
about $1/3$ by using a balance. The target ball is found through an
iterative procedure and the cost is reduced exponentially. By using a
different oracle in each step, the QRT procedure in our algorithm emulates
the usage of the balance in solving the nested structured search problem.

The procedure for solving the nested structured search problem can be
applied for optimization problems that are transformed to finding the ground
state of a problem Hamiltonian in quantum computation. Here, we propose a
quantum algorithm based on multistep quantum computation for optimization
problems with continuous variables. We first discretize the variables of the
objective function to construct the state space of the problem. Then we
construct a sequence of intermediate Hamiltonians to reach the problem
Hamiltonian by decomposing the problem using a set of threshold values, and
apply a multistep quantum computation process to reduce the search space of
the problem step by step. The solution vector to the optimization problem is
narrowed in a small state space and can be determined efficiently through
measurements. If the search spaces of the Hamiltonians are reduced in a
polynomial rate by using an appropriate set of threshold values, then the
optimum of the function can be obtained efficiently. Meanwhile if the global
minimum of the optimization problem is in the state space of the problem,
then it can be obtained efficiently. The problem in many optimization
algorithms where the trial vector is trapped in a deep local minimum and
missing the global minimum can be avoided in our algorithm, provided the
above conditions are satisfied. In quantum computing, the dimension of the
Hilbert space of the qubits increases exponentially with the number of
qubits, it is more efficient to represent a large state space on a quantum
computer than on a classical computer, therefore increasing the probability
of finding the global minimum of the problem.

This paper is organized as follows: in Sec.~II, we describe the quantum
algorithm for optimization problems with continuous variables based on
multistep quantum computation; in Sec.~III, we apply the algorithm for some
test optimization problems, and we close with a discussion.

\section{Quantum optimization algorithm based on multistep quantum
computation}

Let $S$ be the domain of $\mathbf{x}$, an optimization problem can be
formulated as a minimization problem:
\begin{equation}
\text{minimize }F(\mathbf{x})\text{: subject to }\mathbf{x}\in S,
\end{equation}%
where $F$ is a real-valued objective function and $\mathbf{x}$ is the vector
of the variables. Here we focus on optimization problems with continuous
variables, which can be described as follows: for a real-valued function of $%
r$ variables, $F\left( x_{1},x_{2},\cdots ,x_{r}\right) $, find a vector of
the variables such that the function has the minimum value. In the
following, we present a quantum optimization algorithm based on multistep
quantum computation for this problem.

We discretize the continuous variables in the function domain into intervals
of same length for all the variables, and map the problem on a quantum
computer. For simplicity, suppose each variable is discretized into $l$
elements in its definition domain, the dimension of the state space of the
function is $l^{r}$. We prepare $r$ quantum registers and each register
contains $\lceil \log _{2}l\rceil $ qubits that represents the elements of
the variable. Therefore $n=r\lceil \log _{2}l\rceil $ qubits form the
register $R$ that represents the problem with state space of size $N=2^{n}$
on a quantum computer. A vector of the discretized variables $x_{1}^{(i)}$, $%
x_{2}^{(j)}$, $\ldots $, $x_{r}^{(k)}$ is represented by state $|ij\cdots
k\rangle $, where $x_{s}^{(j)}$ represents the $j$th element of the variable
$x_{s}$. The states $|i\rangle $, $|j\rangle $, $\cdots $, $|k\rangle $ are
binary representation of the elements $x_{1}^{(i)}$, $x_{2}^{(j)}$, $\ldots $%
, $x_{r}^{(k)}$ on the quantum registers. These vectors form the
computational basis states~(CBS) of $r$ quantum registers of dimension $N$
as $|J\rangle =|i\rangle |j\rangle \ldots |k\rangle $, $J=0$, $1$, $\ldots $%
, $N-1$, and the corresponding function value is $F\left(
x_{1}^{(i)},x_{2}^{(j)},\ldots ,x_{r}^{(k)}\right) =F\left( J\right) =F_{J}$%
. The task is to find the vector $|Q\rangle =|q_{1}\rangle |q_{2}\rangle
\ldots |q_{r}\rangle $ such that $F\left(
x_{1}^{(q_{1})},x_{2}^{(q_{2})},\ldots ,x_{r}^{(q_{r})}\right) $ is the
minimum of the function $F$.

By using an oracle $O_{F}$\ where $O_{F}|J\rangle |0\rangle =|J\rangle
|F\left( J\right) \rangle $, the Hamiltonian of the optimization problem can
be constructed as%
\begin{equation}
H_{F}|J\rangle =F_{J}|J\rangle ,
\end{equation}%
where $F_{J}$ are eigenvalues of $H_{F}$ with corresponding eigenstates $%
|J\rangle $. The problem of finding the minimum of the function $F$ is
transformed to finding the ground state of the Hamiltonian $H_{F}$ and its
corresponding eigenvalue. We apply a multistep quantum computation process
for solving this problem. We first estimate the range of the function value
as $\left[ F_{\min }\text{, }F_{\max }\right] $, and prepare a set of
threshold values \{$d_{1}$, $d_{2}$, $\ldots $, $d_{m}$\}, and $F_{\max
}>d_{1}>d_{2}>\ldots >d_{m}>F_{\min }$. Then we construct $m$ Hamiltonians
as:%
\begin{equation}
H_{P_{i}}|J\rangle =h_{J}|J\rangle ,\text{ \ \ }i=1,\ldots ,m
\end{equation}%
where
\begin{equation}
h_{J}=\Bigg\lbrace%
\begin{array}{c}
\!\!-1,\,\mathrm{if}\,F_{J}\leq d_{i}, \\
\!\!\hskip.0003in0,\,\mathrm{if}\,F_{J}>d_{i}\,,%
\end{array}%
\end{equation}%
and $H_{P_{m}}=H_{m}=H_{P}$. This can be achieved by using an oracle that
recognizes whether $F_{J}$ is larger or less than a threshold value $d_{i}$.
It is a comparison logic circuitry and can be implemented efficiently on a
quantum computer~\cite[p.264]{nc}~\cite{durr, bari, grandunif}. The CBS
associated with integers that are less than or equal to $d_{i}$ form a set $%
A_{i}$ with size $N_{i}$. They have the nested structure as $A_{m}\subset
A_{m-1}\subset \cdots \subset A_{1}$. The ground state of the problem
Hamiltonian $H_{P}$ contains CBS in $A_{m}$ with eigenvalues that are below
the threshold value $d_{m}$. We construct a sequence of Hamiltonians that
form a Hamiltonian evolution path to the problem Hamiltonian $H_{P}$ as%
\begin{equation}
H_{i}=\frac{M_{i}}{N}H_{0}+\left( 1-\frac{M_{i}}{N}\right) H_{P_{i}},\text{
\ \ }i=0,1,\ldots ,m-1,
\end{equation}%
where $H_{0}=-|\psi _{0}\rangle \langle \psi _{0}|$ and $|\psi _{0}\rangle =%
\frac{1}{\sqrt{N}}\sum_{j=0}^{N-1}|j\rangle $, and $M_{i}$ is an approximate
estimation of $N_{i}$. We have demonstrated that as $M_{i}=N_{i}$, the
conditions for efficiently running the algorithm are satisfied provided that
the ratio $N_{i-1}/N_{i}$ are polynomial large~\cite{wyx}. The parameters $%
M_{i}$ can be estimated efficiently by using the Monte Carlo sampling method~%
\cite{wanglandau}, and we can adjust the threshold values such that the
ratio $N_{i-1}/N_{i}$ are polynomial large. Detailed analysis of the effect
of $M_{i}$ on the efficiency of the algorithm is presented in the appendix.
The ground state of $H_{P}$ can be obtained through the following multistep
quantum computation process based on QRT in $m$ steps. We use the $i$th step
of the algorithm to illustrate the procedures.

In the $i$th step, given the Hamiltonian $H_{i-1}$, its ground state
eigenvalue $E_{0}^{\left( i-1\right) }$ and the ground state $|\varphi
_{0}^{\left( i-1\right) }\rangle $ obtained from the previous step, we are
to prepare the ground state $|\varphi _{0}^{\left( i\right) }\rangle $ of $%
H_{i}$ by using the QRT method. The algorithm requires ($n+1$) qubits with a
probe qubit coupling to the $n$-qubit register $R$. The algorithm
Hamiltonian of the $i$th step is constructed as
\begin{equation}
H^{\left( i\right) }=-\frac{1}{2}\omega \sigma _{z}\otimes
I_{N}+H_{R}^{\left( i\right) }+c\sigma _{x}\otimes I_{N},
\end{equation}%
where
\begin{equation}
H_{R}^{\left( i\right) }=\alpha _{i}|1\rangle \langle 1|\otimes
H_{i-1}+|0\rangle \langle 0|\otimes H_{i}\mathbf{,}\text{\ }i=1,2,\cdots ,m,
\end{equation}%
$I_{N}$ is the $N$-dimensional identity operator, and $\sigma _{x}$ and $%
\sigma _{z}$ are the Pauli matrices. The first term in Eq.~($9$) is the
Hamiltonian of the probe qubit, the second term contains the Hamiltonian of
the register $R$ and describes the interaction between the probe qubit and $%
R $, and the third term is a perturbation with $c\ll 1$. The parameter $%
\alpha _{i}$ is used to re-scale the energy levels of $H_{i-1}$, and the
ground state energy of $\alpha _{i}H_{i-1}$ is used as a reference energy
level to the ground state eigenvalue $E_{0}^{\left( i\right) }$ of $H_{i}$.
The initial state of the $(n+1)$ qubits is set as $|1\rangle |\varphi
_{0}^{\left( i-1\right) }\rangle $, which is an eigenstate of $H_{R}^{\left(
i\right) }$ with eigenvalue $\alpha _{i}E_{0}^{\left( i-1\right) }$. First
we obtain the eigenvalue $E_{0}^{\left( i\right) }$\ of $H_{i}$ by using the
QRT method through varying the frequency of the probe qubit as shown in Ref.~%
\cite{wyx}. Then we set $\alpha _{i}=\left( E_{0}^{\left( i\right) }-\omega
\right) /E_{0}^{\left( i-1\right) }$, such that the condition of $%
E_{0}^{\left( i\right) }-\alpha _{i}E_{0}^{\left( i-1\right) }=\omega $ for
resonant transition between the probe qubit and the transition between
states $|\varphi _{0}^{\left( i-1\right) }\rangle $ and $|\varphi
_{0}^{\left( i\right) }\rangle $ is satisfied. When obtaining the eigenvalue
$E_{0}^{\left( i\right) }$\ of $H_{i}$, we can also obtain the overlap $%
g_{0}^{(i)}=\langle \varphi _{0}^{(i-1)}|\varphi _{0}^{(i)}\rangle $ between
the ground states of $H_{i-1}$ and $H_{i}$ through the Rabi's formula~\cite%
{cohen}. Then we can set the optimal runtime $t_{i}=\pi /(2cg_{0}^{(i)})$ at
which the probability for the system to be evolved to the state $|0\rangle
|\varphi _{0}^{\left( i\right) }\rangle $ reaches its maximum. The
procedures for obtaining the ground state of $H_{i}$ are summarized as
follows:

($i$) Initialize the probe qubit to its excited state $|1\rangle $ and the
register $R$ in state $|\varphi _{0}^{\left( i-1\right) }\rangle $;

($ii$) Implement the unitary evolution operator $U(t_{i})=\exp \left(
-iH^{\left( i\right) }t_{i}\right) $;

($iii$) Read out the state of the probe qubit.

The system is approximately in state $\sqrt{1-p_{0}^{\left( i\right) }}%
|1\rangle |\varphi _{0}^{\left( i-1\right) }\rangle +\sqrt{p_{0}^{\left(
i\right) }}|0\rangle |\varphi _{0}^{\left( i\right) }\rangle $ as the
resonant transition occurs, where $p_{0}^{\left( i\right) }=\sin ^{2}\left(
ct_{i}g_{0}^{(i)}\right) $ is the decay probability of the probe qubit of
the $i$th step. The state $|\varphi _{0}^{\left( i-1\right) }\rangle $ from
the previous step is protected in this entangled state. By performing a
measurement on the probe qubit, if the probe decays to its ground state $%
|0\rangle $, it indicates that the resonant transition occurs and the system
evolves from the state $|1\rangle |\varphi _{0}^{\left( i-1\right) }\rangle $
to the state $|0\rangle |\varphi _{0}^{\left( i\right) }\rangle $; otherwise
if the probe qubit stays in state $|1\rangle $, it means that the register $%
R $ remains in state $|\varphi _{0}^{\left( i-1\right) }\rangle $, then we
repeat procedures $ii$)-$iii$) until the probe qubit decays to its ground
state $|0\rangle $. Therefore we can obtain the ground state $|\varphi
_{0}^{\left( i\right) }\rangle $ of $H_{i}$ deterministically. By protecting
the state $|\varphi _{0}^{\left( i-1\right) }\rangle $ through entanglement,
the state can be used repeatedly without copying it, such that the algorithm
realizes multistep quantum computation. The runtime of the algorithm is
proportional to the number of steps of the algorithm, and the success
probability of the algorithm is polynomial large by setting the coupling
coefficient $c$ appropriately~\cite{wyx}. After running the algorithm for $m$
steps, we obtain the ground state of the problem Hamiltonian $H_{P}$, which
is a superposition state of a few CBS with eigenvalues below the threshold
value $d_{m}$. Then we can perform measurement on the state and find the CBS
that has the minimum function value, therefore solving the optimization
problem. We can run the algorithm for a few rounds by discretizing the
variables in the neighborhood of the optimized vector to improve the
precision of the solution to the optimization problem.

The algorithm can be run efficiently if both the energy gap between the
ground and the first excited states of each Hamiltonian $H_{i}$ and the
overlap between ground states of any two adjacent Hamiltonians $g_{0}^{(i)}$
are not exponentially small. By solving the eigen-problem of the Hamiltonian
$H_{i}$, these conditions can be satisfied if the ratio $N_{i-1}/N_{i}$ are
polynomial large, and the parameters $M_{i}$ are set such that: the point $%
\left( N_{i}/N,M_{i}/N\right) $ is far away from the neighborhood of the
point $\left( 0,1/2\right) $, and $2M_{i}\left( N-N_{i}\right) <N^{2}$.
Detailed analysis is shown in the appendix.

\section{Application of the algorithm for some test functions}

We now apply the quantum optimization algorithm described above for some
test functions of optimization problem: the Damavandi function, the Griewank
function and the Price function.

\subsection{The Damavandi function}

The two dimensional Damavandi function is defined as
\begin{widetext}
\begin{equation}
f_{\text{Damavandi}}\left( x_{1},x_{2}\right) =\left[ 1-\left\vert \frac{%
\sin \left[ \pi \left( x_{1}-2\right) \right] \sin \left[ \pi \left(
x_{2}-2\right) \right] }{\pi ^{2}\left( x_{1}-2\right) \left( x_{2}-2\right)
}\right\vert ^{5}\right] \left[ 2+\left( x_{1}-7\right) ^{2}+2\left(
x_{2}-7\right) ^{2}\right] ,
\end{equation}
\end{widetext}
and the graph of this function is shown in Fig.~$1$. It has a very sharp
global minimum of zero at \{$x_{1}=2$, $x_{2}=2$\}. For classical
optimization algorithms based on the gradients methods, it is very easy for
a trial vector to be trapped in the bowl-like local minimum, while missing
the global minimum. The overall success probability of current global
optimization algorithms for finding the global minimum of this function is
about $0.25\%$~\cite{testfunc}.

To apply our algorithm for this optimization problem, the two variables of
the Damavandi function are discretized into $281$ elements evenly with an
interval of $0.05$ in the range $\left[ 0,14\right] $. The dimension of the
state space of the function is $78961$. By discretizing the value of the
function in an interval of $0.01$ in the range of $\left[ 0,149\right] $,
and counting the number of states in each interval, we can obtain the
distribution of states of the function in each energy interval as shown in
Fig.~$2$. The largest degeneracy is about $56$, which is a small number
compare to the dimension of the state space. We construct a set of threshold
values as \{$d_{1}=70$, $d_{2}=30$, $d_{3}=15$, $d_{4}=7$, $d_{5}=4$, $%
d_{6}=3$, $d_{7}=2.5$, $d_{8}=2.2$, $d_{9}=2.1$, $d_{10}=2.02$, $d_{11}=2.0$%
, $d_{12}=1.0$\}, and run the algorithm. The dimension of the corresponding
state space in each step of the algorithm are reduced to \{$56634$, $24939$,
$11573$, $4452$, $1772$, $892$, $448$, $178$, $94$, $20$, $5$, $1$\},
respectively. The dimension of the state space is reduced smoothly with
reduction rates of \{$0.717$, $0.440$, $0.464$, $0.385$, $0.398$, $0.503$, $%
0.502$, $0.397$, $0.528$, $0.213$, $0.250$, $0.200$\} in each step of the
algorithm, respectively. The parameter $M_{i}$ can be estimated through
Monte Carlo sampling, if it is set approximately as $N_{i}$ above, the
conditions of the algorithm can be satisfied. The ratio $M_{i}/N$ that are
closest to $1/2$ is $0.316$.

We can see that the dimension of the state space is reduced to a few CBS
after a number of steps. Therefore the final state that encodes the solution
to the optimization problem can be readout and checked efficiently to find
the global optimum of the function.
\begin{figure}[tb]
\includegraphics[width=0.83\columnwidth, clip]{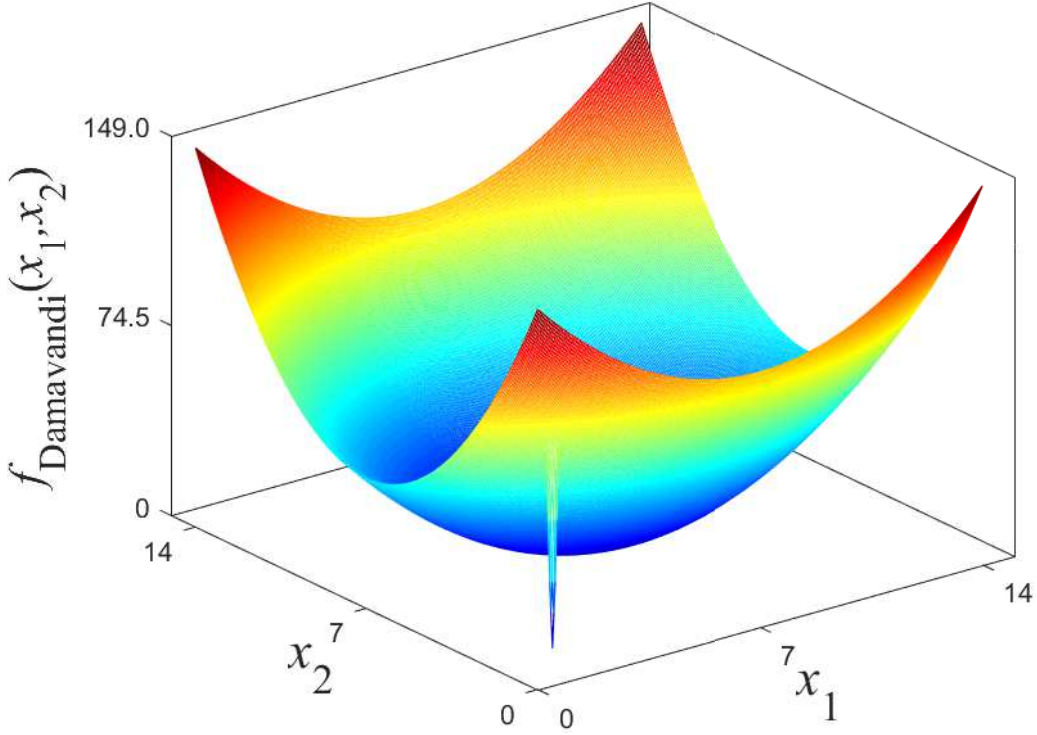}
\caption{ (Color online)~The graph of the Damavandi function.}
\end{figure}
\begin{figure}[tb]
\includegraphics[width=0.83\columnwidth, clip]{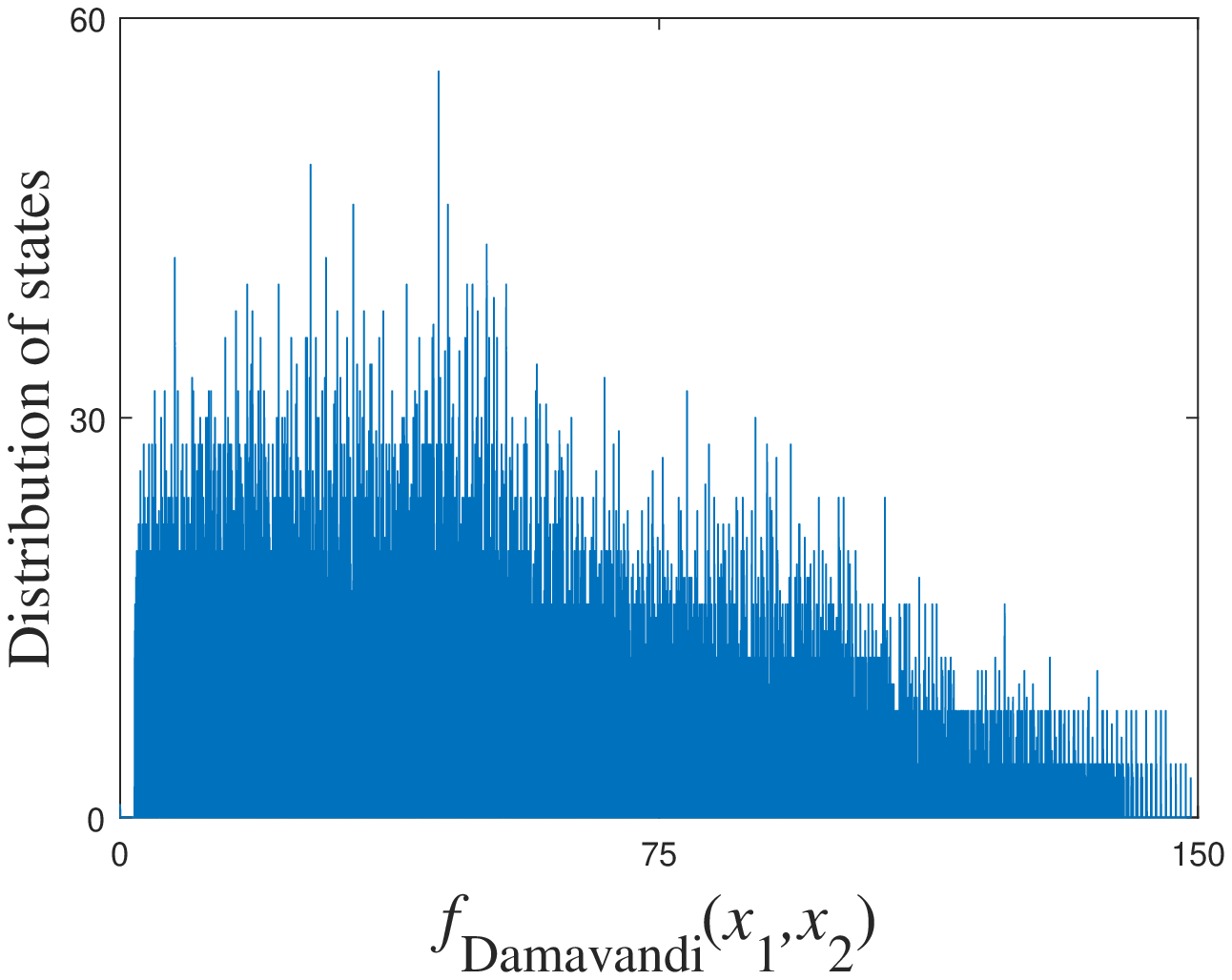}
\caption{Distribution of states by discretizing the value of the Damavandi
function in intervals of $0.01$.}
\end{figure}

\subsection{The Griewank function}

The Griewank function has the form
\begin{equation}
f_{\text{Griewank}}\!\left(\! x_{1},\cdots ,x_{n}\!\right) \!=\!\frac{1}{4000%
}\sum_{k=1}^{n}\!x_{k}^{2}\!-\!\prod_{k=1}^{n}\! \cos\! \left(\!\frac{x_{k}}{%
\sqrt{k}}\!\right) \!+\! 1.
\end{equation}
Fig.~$3$ shows the second-order Griewank function with two variables, we can
see that the function has many local minima. For classical optimization
algorithms, it is very easy for a trial vector to be trapped in one of the
local minima, while missing the global minimum of the function. This
situation can be avoided in our algorithm.

We discretize the two variables of the Griewank function into $801$ elements
evenly with interval of $0.1$ in the range $\left[ -40,40\right] $. The
dimension of the state space of the function is $641601$. By discretizing
the function value in intervals of $0.0001$, the distribution of states in
each energy interval of the function is shown in Fig.~$4$. The largest
degeneracy is $32$. The threshold value set is constructed as \{$d_{1}=1.0$,
$d_{2}=0.6$, $d_{3}=0.4$, $d_{4}=0.3$, $d_{5}=0.2$, $d_{6}=0.1$, $d_{7}=0.06$%
, $d_{8}=0.04$, $d_{9}=0.02$, $d_{10}=0.01$, $d_{11}=0.005$, $d_{12}=0.002$%
\}. The sizes of the corresponding state spaces for each step of the
algorithm are \{$197363$, $76951$, $34453$, $18937$, $8283$, $2033$, $723$, $%
319$, $77$, $23$, $5$, $1$\}, respectively. The dimension of the state space
of the problem is reduced smoothly in each step of the algorithm in rate of
\{$0.31$, $0.39$, $0.45$, $0.55$, $0.44$, $0.25$, $0.36$, $0.44$, $0.24$, $%
0.30$, $0.22$, $0.20$\}, respectively. After running the algorithm for a
number of steps, the state space of the problem is reduced to a very small
space and can be readout to calculate the corresponding function value and
find the global minimum.
\begin{figure}[tb]
\includegraphics[width=0.83\columnwidth, clip]{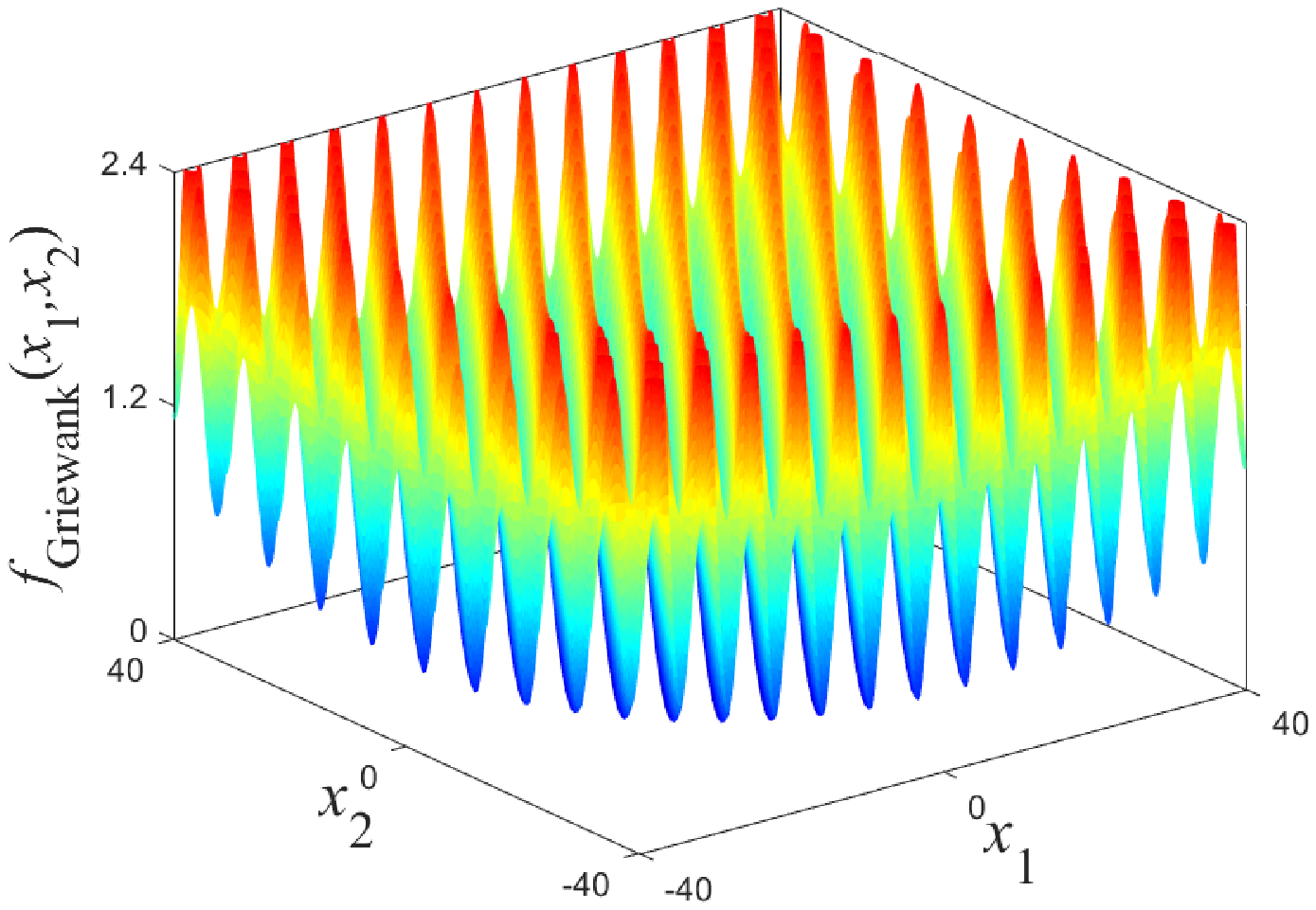}
\caption{ (Color online)~The second-order Griewank function.}
\end{figure}
\begin{figure}[tb]
\includegraphics[width=0.83\columnwidth, clip]{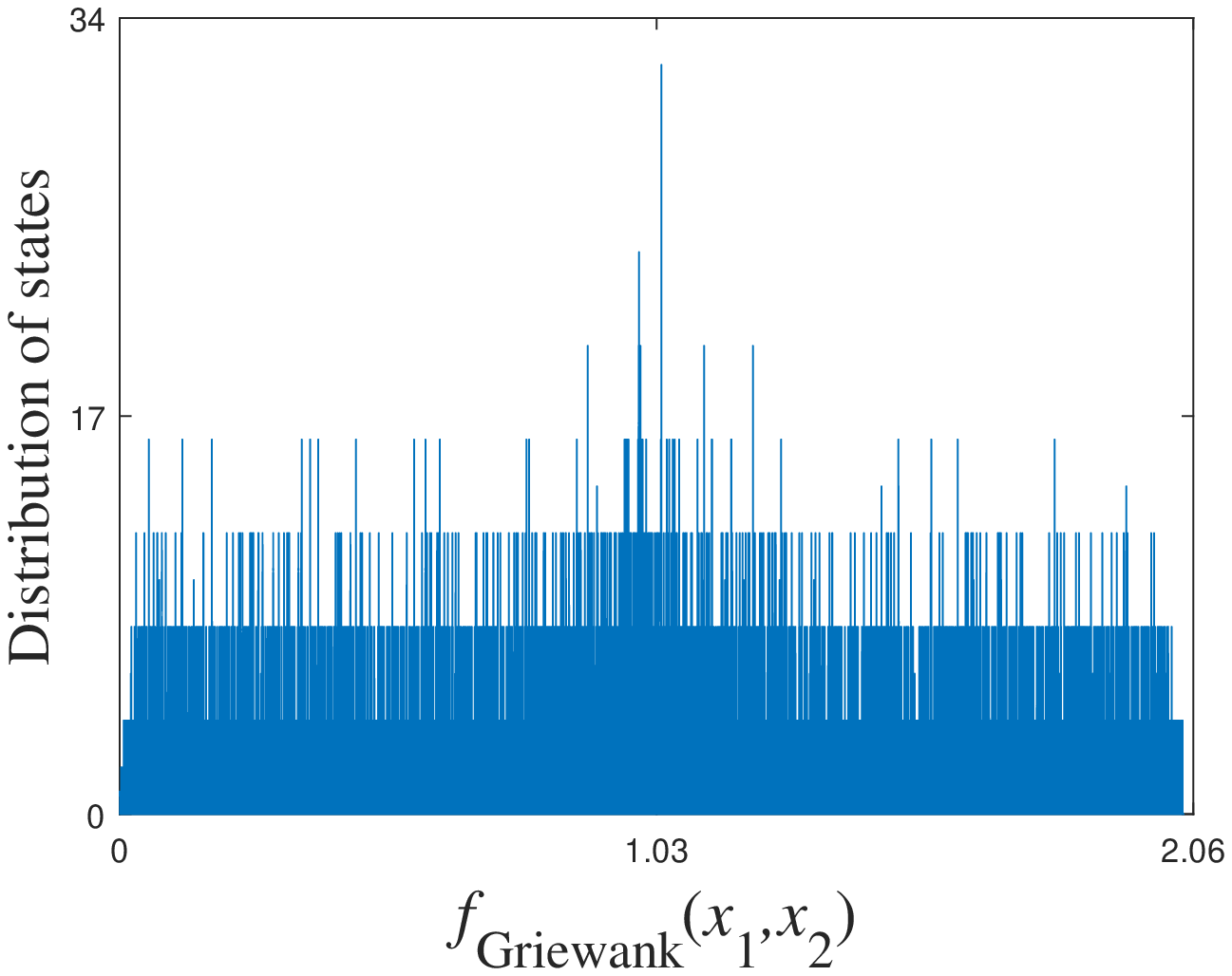}
\caption{Distribution of states by discretizing the value of the
second-order Griewank function in intervals of $0.0001$.}
\end{figure}

\subsection{The Price function}

The Price$01$ function can be written in form of%
\begin{equation}
f_{\text{Price}}\left( x_{1},x_{2}\right) =\left( \left\vert
x_{1}\right\vert -5\right) ^{2}+\left( \left\vert x_{2}\right\vert -5\right)
^{2},
\end{equation}%
with four minima as shown in Fig.~$5$. Our algorithm can be applied to
obtain the four vectors corresponding to the minimum of the function in a
degenerate state.
\begin{figure}[tb]
\includegraphics[width=0.83\columnwidth, clip]{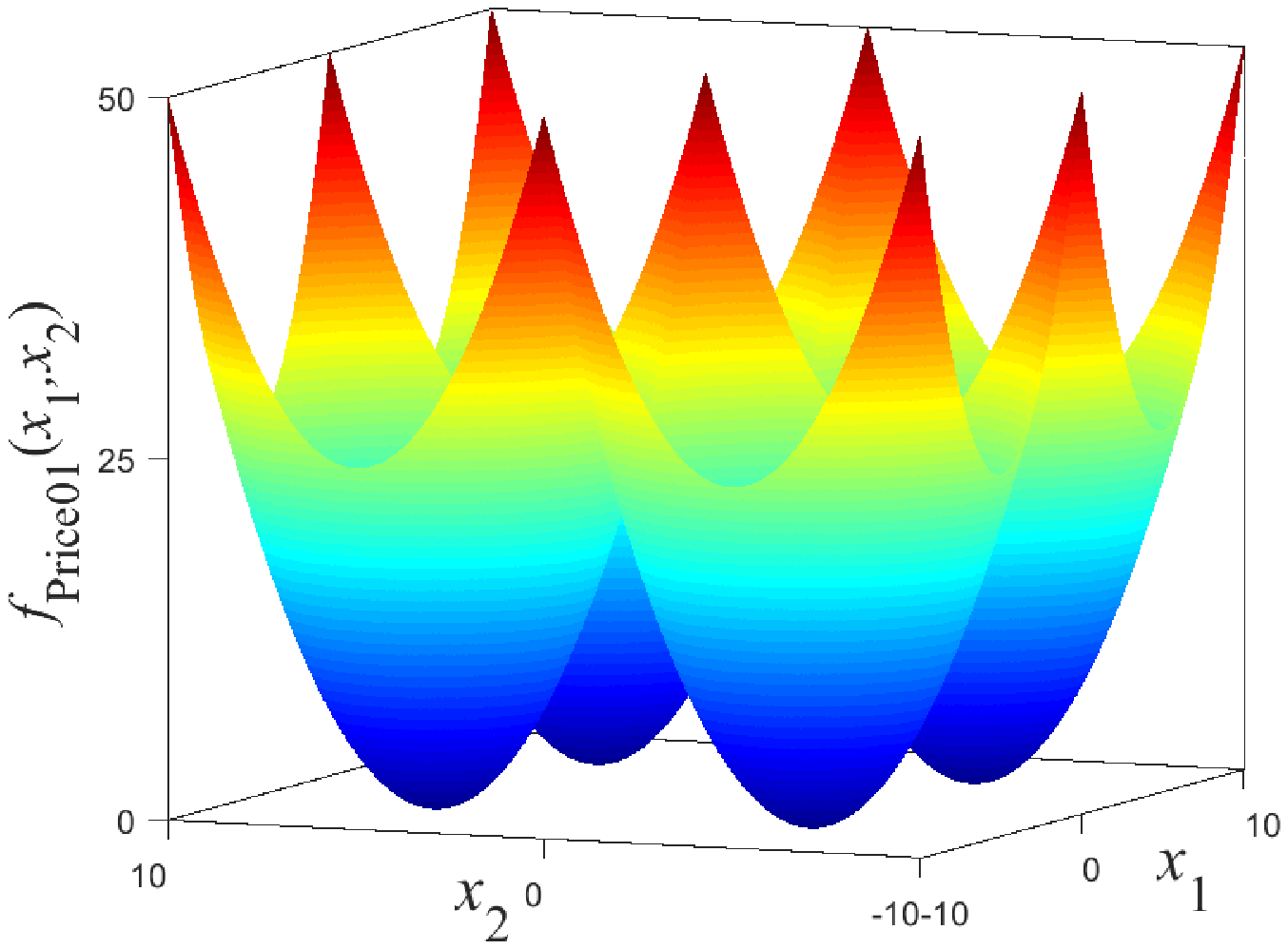}
\caption{ (Color online)~The Price$01$ function.}
\end{figure}

The two variables of the Price function are discretized into $201$ elements
evenly with an interval of $0.1$ in the range $\left[ -10,10\right] $. The
dimension of the state space of the problem is $40401$. A set of threshold
values is constructed as \{$d_{1}=20$, $d_{2}=10$, $d_{3}=5$, $d_{4}=2$, $%
d_{5}=1$, $d_{6}=0.5$, $d_{7}=0.2$, $d_{8}=0.1$, $d_{9}=0.05$, $d_{10}=0.02$%
, $d_{11}=0.01$\}. The dimension of the corresponding state space in each
round of the algorithm are reduced to \{$25108$, $12532$, $6260$, $2484$, $%
1220$, $596$, $244$, $116$, $52$, $20$, $4$\}, respectively. The
corresponding reduction rates in each round of the algorithm are \{$0.62$, $%
0.50$, $0.50$, $0.40$, $0.49$, $0.49$, $0.41$, $0.48$, $0.45$, $0.38$, $0.20$%
\}. The final state is in an equal superposition of the four global minima
of the function and can be obtained by readout of the state of the circuit.

\section{Discussion}

In this work, we present a quantum optimization algorithm for solving
optimization problems with continuous variables based on multistep quantum
computation. The state space of the problem is constructed by discretizing
the variables of the objective function. By applying a multistep quantum
computation process, the search space of the problem can be reduced step by
step. We construct a sequence of Hamiltonians based on a set of threshold
values, such that the search spaces corresponding to the Hamiltonians form a
nested structure. If the dimension of search spaces is reduced sequentially
in polynomial rate, then the algorithm can be run efficiently. The reduction
rate can be adjusted by setting the threshold values appropriately. The
final state obtained by the algorithm is a superposition of a few CBS~(or a
CBS) and the minimum of the function can be determined efficiently by
measuring the state and evaluating the corresponding function value.

One of the most difficult problems for optimization algorithms is that a
trial vector is trapped in a deep local minimum, while missing the global
minimum. In our algorithm, we locate the global minimum of the problem by
using a number of threshold values, and obtain the corresponding state
vector through a multistep quantum computation process by narrowing the
search space of the problem step by step. The global minimum can be obtained
if it is in the state space of the problem and the conditions of the
algorithm are satisfied. One advantage of quantum computing is that
exponential number of CBS can be stored in polynomial number of qubits.
Therefore we can construct a large state space of the problem by using a
small number of qubits, such that increasing the probability of finding the
global minimum of the objective function. The precision of the algorithm can
be improved by running the algorithm for a few rounds in the neighborhood of
the minimum being found.

\begin{acknowledgements}
We thank A. Miranowicz and F. Nori for helpful discussions. This work was supported by National Key Research and Development Program of China~(2021YFA1000600), the Fundamental Research Funds for the Central Universities~(Grant No.~11913291000022), and the Natural Science Fundamental Research Program of Shaanxi Province of China under grant No.~2022JM-021.
\end{acknowledgements}

\appendix

\begin{appendix}

\section{Solving the eigen-problem of the intermediate Hamiltonians}

In the following, we solve the eigen-problem of the intermediate Hamiltonian
to calculate the energy gap between the ground and the first excited states
of the Hamiltonian, and the overlap between the ground states of two
adjacent Hamiltonians.

In the quantum optimization algorithm, we construct a sequence of
intermediate Hamiltonians to form a Hamiltonian evolution path to the
problem Hamiltonian as
\begin{equation}
H_{i}=\frac{M_{i}}{N}H_{0}+\left( 1-\frac{M_{i}}{N}\right) H_{P_{i}},\text{
\ }i=1,2,\cdots ,m
\end{equation}%
where%
\begin{equation}
H\!_{0}\!\!=-|\psi _{0}\rangle \langle \psi _{0}|,
\end{equation}%
with $|\psi _{0}\rangle =\frac{1}{\sqrt{N}}\sum_{j=0}^{N-1}|j\rangle $, and
\begin{equation}
H_{P_{i}}=-\sum_{q_{i}\in A_{i}}|q_{i}\rangle \langle q_{i}|,
\end{equation}%
where $A_{1}\supset \cdots \supset A_{m-1}\supset A_{m}$ are determined by
Eqs.~($6$) and ($7$), and the sizes of the sets $A_{1}$, $\cdots $, $A_{m}$
are $N_{1}$, $\cdots $, $N_{m}$, respectively, and $N_{1}>\cdots >N_{m}$.
Let $H_{m}=H_{P}=H_{P_{m}}$ and the set $A_{m}$ contains the target states $%
|q\rangle $ with size $N_{m}$. We construct a Hamiltonian evolution path $%
H_{0}\rightarrow H_{1}\rightarrow \cdots \rightarrow H_{m}=H_{P}$ and start
from the ground state $|\varphi _{0}^{(0)}\rangle $ of $H_{0}$, evolve it
through ground states of the intermediate Hamiltonians sequentially via
quantum resonant transition~(QRT), finally reach the ground state $|\varphi
_{0}^{(m)}\rangle$ of $H_{P}$ in $m$ steps. The algorithm can be run
efficiently provided: ($i$) the energy gap between the ground and the first
excited states of each Hamiltonian and, ($ii$) the overlaps between ground
states of any two adjacent Hamiltonians are not exponentially small.

In the following we solve the eigen-problem of the Hamiltonian $H_{i}$. Let
\begin{equation}
|\psi _{0}\rangle =\frac{1}{\sqrt{N}}\sum_{j=0}^{N-1}|j\rangle =\frac{1}{%
\sqrt{N}}\sum_{q_{i}\in A_{i}}|q_{i}\rangle +\sqrt{\frac{N-N_{i}}{N}}%
|q_{i}^{\bot }\rangle .
\end{equation}%
where%
\begin{equation}
|q_{i}^{\bot }\rangle =\frac{1}{\sqrt{N-N_{i}}}\sum_{k\notin A_{i}}|k\rangle
.
\end{equation}%
Then in basis $\left( \left\{ |q_{i}\rangle \right\} _{q_{i}\in A_{i}}\text{%
, }|q_{i}^{\bot }\rangle \right) $, we have%
\begin{eqnarray}
H\!_{0}\! &=&-|\psi _{0}\rangle \langle \psi _{0}|  \notag \\
&=&-\left( \!%
\begin{array}{cccc}
\frac{1}{N} & \cdots & \frac{1}{N} & \frac{\sqrt{\!N\!-\!N_{i}}}{N} \\
\vdots & \ddots & \vdots & \vdots \\
\frac{1}{N} & \cdots & \frac{1}{N} & \frac{\sqrt{N\!-\!N_{i}}}{N} \\
\frac{\sqrt{\!N\!-\!N_{i}}}{N} & \cdots & \frac{\sqrt{\!N\!-\!N_{i}}}{N} &
\frac{\!N\!-\!N_{i}\!}{N}\!%
\end{array}%
\!\right) ,
\end{eqnarray}
and
\begin{eqnarray}
H_{P_{i}} &=&-\sum_{q_{i}\in A_{i}}|q_{i}\rangle \langle q_{i}|  \notag \\
&=&-\!\left( \!%
\begin{array}{cccc}
1 & \cdots & 0 & 0 \\
\vdots & \ddots & \vdots & \vdots \\
0 & \cdots & 1 & 0 \\
0 & \cdots & 0 & 0%
\end{array}%
\!\right) .
\end{eqnarray}%
Then
\begin{widetext}
\begin{eqnarray}
H_{i} &=&\frac{M_{i}}{N}H_{0}+\left( 1-\frac{M_{i}}{N}\right) H_{P_{i}}
\notag \\
&=&-\!\!\left( \!%
\begin{array}{ccccc}
\frac{M_{i}}{N^{2}}+1-\frac{M_{i}}{N} & \frac{M_{i}}{N^{2}} & \cdots & \frac{%
M_{i}}{N^{2}} & \frac{M_{i}\sqrt{\!N\!-\!N_{i}}}{N^{2}} \\
\frac{M_{i}}{N^{2}} & \frac{M_{i}}{N^{2}}+1-\frac{M_{i}}{N} & \cdots & \vdots
& \vdots \\
\vdots & \vdots & \ddots &  &  \\
\frac{M_{i}}{N^{2}} & \frac{M_{i}}{N^{2}} & \cdots & \frac{M_{i}}{N^{2}}+1-%
\frac{M_{i}}{N} & \frac{M_{i}\sqrt{N\!-\!N_{i}}}{N^{2}} \\
\frac{M_{i}\sqrt{\!N\!-\!N_{i}}}{N^{2}} & \frac{M_{i}\sqrt{\!N\!-\!N_{i}}}{%
N^{2}} & \cdots & \frac{M_{i}\sqrt{\!N\!-\!N_{i}}}{N^{2}} & \frac{%
M_{i}\left( N\!-\!N_{i}\right) \!\!}{N^{2}}%
\end{array}%
\!\right) .
\end{eqnarray}
\end{widetext}

Let $n=N_{i}+1$,
\begin{equation}
\mathbf{e}=\left(
\begin{array}{cccc}
1 & \cdots  & 1 & 1%
\end{array}%
\right) _{1\times n}^{T},\qquad \mathbf{e}_{n}=\left(
\begin{array}{cccc}
0 & \cdots  & 0 & 1%
\end{array}%
\right) _{1\times n}^{T},  \label{Definitions_e_en}
\end{equation}%
then $H_{0}$ can be rewritten as
\begin{widetext}
\begin{eqnarray}
H_{0} &=&-\left( \frac{1}{\sqrt{N}}\mathbf{e}+\frac{\sqrt{N-N_{i}}-1}{\sqrt{N%
}}\mathbf{e}_{n}\right) \left( \frac{1}{\sqrt{N}}\mathbf{e}+\frac{\sqrt{%
N-N_{i}}-1}{\sqrt{N}}\mathbf{e}_{n}\right) ^{T}  \notag \\
&=&-\frac{1}{N}\mathbf{e}\mathbf{e}^{T}-\frac{\sqrt{N-N_{i}}-1}{N}\left(
\mathbf{e}\mathbf{e}_{n}^{T}+\mathbf{e}_{n}\mathbf{e}^{T}\right) -\frac{%
\left( \sqrt{\!N\!-\!N_{i}}-1\right) ^{2}}{N}\mathbf{e}_{n}\mathbf{e}%
_{n}^{T},
\end{eqnarray}
\end{widetext}and%
\begin{equation}
H_{P_{i}}=-\left( I_{n}-\mathbf{e}_{n}\mathbf{e}_{n}^{T}\right) ,
\end{equation}%
where $I_{n}$ is the $(N_{i}+1)$-dimensional identity operator. Thus
\begin{widetext}
\begin{eqnarray}
H_{i} &=&\frac{M_{i}}{N}H_{0}+\left( 1-\frac{M_{i}}{N}\right) H_{P_{i}}
\notag \\
&=&\!-\frac{M_{i}}{N}\!\left[ \!\frac{1}{N}\mathbf{e}\mathbf{e}^{T}\!+\!%
\frac{\sqrt{N\!-\!N_{i}}\!-\!1}{N}\left( \!\mathbf{e}\mathbf{e}_{n}^{T}\!+\!%
\mathbf{e}_{n}\mathbf{e}^{T}\!\right) \!+\!\frac{\left( \!\sqrt{\!N\!-\!N_{i}%
}\!-\!1\!\right) ^{2}}{N}\mathbf{e}_{n}\mathbf{e}_{n}^{T}\!\right]
\!\!-\!\!\left( \!1\!-\!\frac{M_{i}}{N}\!\right) \left( \!I_{n}\!-\!\mathbf{e%
}_{n}\mathbf{e}_{n}^{T}\!\right)  \notag \\
&=&-\frac{M_{i}}{N^{2}}\mathbf{e}\mathbf{e}^{T}\!-\!\frac{M_{i}\left( \!%
\sqrt{N-N_{i}}\!-\!1\!\right) }{N^{2}}\left( \mathbf{e}\mathbf{e}%
_{n}^{T}\!+\!\mathbf{e}_{n}\mathbf{e}^{T}\right) \!+\!\left[ \!\left(
\!1\!-\!\frac{M_{i}}{N}\!\right) \!-\!\frac{M_{i}\left( \!\sqrt{\!N\!-\!N_{i}%
}\!-\!1\!\right) ^{2}}{N^{2}}\!\right] \mathbf{e}_{n}\mathbf{e}_{n}^{T}
\notag \\
&&-\left( 1-\frac{M_{i}}{N}\right) I_{n}  \notag \\
&\equiv &\alpha \mathbf{e}\mathbf{e}^{T}+\beta \left( \mathbf{e}\mathbf{e}%
_{n}^{T}+\mathbf{e}_{n}\mathbf{e}^{T}\right) +\left( \gamma -2\beta \right)
\mathbf{e}_{n}\mathbf{e}_{n}^{T}-\left( 1-\frac{M_{i}}{N}\right) I_{n},
\label{formula_intermediateHj}
\end{eqnarray}
\end{widetext}where $\alpha =-\frac{M_{i}}{N^{2}}$, $\beta =-\frac{%
M_{i}\left( \sqrt{N-N_{i}}-1\right) }{N^{2}}$, $\gamma =2\beta +1-\frac{M_{i}%
}{N}-\frac{M_{i}\left( \sqrt{\!N\!-\!N_{i}}-1\right) ^{2}}{N^{2}}$. Please
note that, with a bit abuse of notation, in the following we will reuse the
notations $\mathbf{e}$, $\mathbf{e}_{n}$, $\alpha $, $\beta $ and $\gamma $,
and their dimensions and values can be determined easily from the context.

Define $\tilde{\mathbf{e}}$ to be an $N_{i}\times 1$ vector of all ones and
we can rewrite $H_{i}$ as
\begin{eqnarray}
H_{i} &=&\alpha \mathbf{e}\mathbf{e}^{T}\!+\!\left(
\begin{array}{cc}
\mathbf{0}_{N_{i}} & \beta \tilde{\mathbf{e}} \\
\beta \tilde{\mathbf{e}}^{T} & \gamma%
\end{array}%
\right) -\left( 1-\frac{M_{i}}{N}\right) I_{n}  \notag \\
&\equiv &\alpha \mathbf{e}\mathbf{e}^{T}+G-\left( 1-\frac{M_{i}}{N}\right)
I_{n}.
\end{eqnarray}

($i$) Define the vector space $V=\left\{ \mathbf{e},\mathbf{e}_{n}\right\} $
of dimension $2$. Then from Eq.~\eqref{formula_intermediateHj}, $\forall
x\in V^{\perp }$, we have $H_{i}x=-\left( 1-\frac{M_{i}}{N}\right) x$,
therefore the eigenvalues are $-\left( 1-\frac{M_{i}}{N}\right) $,
corresponding to $N_{i}-1$ eigenvectors.

($ii$) The vector space of $V$ can be spanned by vectors%
\begin{equation}
W=\left[ \frac{1}{\sqrt{N_{i}}}\left(
\begin{array}{c}
1 \\
\vdots \\
1 \\
0%
\end{array}%
\right) ,\left(
\begin{array}{c}
0 \\
\vdots \\
0 \\
1%
\end{array}%
\right) \right] .
\end{equation}%
It is easy to check that%
\begin{equation}
W^{T}GW=\left(
\begin{array}{cc}
0 & \beta \sqrt{N_{i}} \\
\beta \sqrt{N_{i}} & \gamma%
\end{array}%
\right) ,
\end{equation}%
and
\begin{equation}
W^{T}\mathbf{e}=\left(
\begin{array}{c}
\sqrt{N_{i}} \\
1%
\end{array}%
\right) .
\end{equation}%
Then we can verify that%
\begin{eqnarray}
W^{T}H_{i}W &=&\alpha W^{T}\mathbf{e}\mathbf{e}^{T}W+W^{T}GW-\left( 1-\frac{%
M_{i}}{N}\right)\! I  \notag \\
&=&\left(\!\!
\begin{array}{cc}
\!\alpha N_{i} & \left(\! \alpha \!+\! \beta \!\right)\! \sqrt{N_{i}} \\
\left(\! \alpha \!+\!\beta \! \right)\! \sqrt{\!N_{i}} & \alpha
\!\!+\!\!\gamma%
\end{array}%
\!\!\right) \!\!-\!\!\left(\!\! 1\!\!-\!\! \frac{M_{i}}{N}\!\!\right)\!I.
\label{formulaHj4SinglePoint}
\end{eqnarray}
Solving the eigen-problem of the above matrix, we can obtain the
eigenvalues:
\begin{widetext}
\begin{eqnarray}
E_{\pm }\! &=&\frac{1}{2}\left( \alpha N_{i}\!+\!\alpha \!+\!\gamma \right)
\pm \frac{1}{2}\sqrt{(\alpha N_{i}\!+\!\alpha \!+\!\gamma )^{2}\!+\!4(\alpha
\!+\!\beta )^{2}N_{i}\!-\!4\alpha N_{i}(\alpha \!+\!\gamma )}\!-\!\left(
1\!-\!\frac{M_{i}}{N}\right)  \notag \\
&=&-\frac{1}{2}\pm \frac{1}{2}\sqrt{1-4\frac{M_{i}}{N}+4\frac{M_{i}N_{i}}{%
N^{2}}+4\left( \frac{M_{i}}{N}\right) ^{2}-4\frac{N_{i}M_{i}^{2}}{N^{3}}}.
\label{eqn:lambda4IntermediateHj}
\end{eqnarray}
\end{widetext}
Besides these $N_{i}+1$ eigenvalues above, there are also $N-(N_{i}+1)$
degenerate eigenstates with eigenvalue $0$, and they are orthogonal to both
the vector space $V=\left\{ \mathbf{e},\mathbf{e}_{n}\right\} $ and the
vector space $V^{\perp }$ of dimension $N_{i}-1$.

In the following, we evaluate the energy gap between the ground and the
first excited states of the intermediate Hamiltonians, and the overlap
between the ground states of two adjacent intermediate Hamiltonians, to
figure out how to satisfy the conditions of the algorithm.

($i$) Estimation of the energy gap between the ground and the first excited
states of the intermediate Hamiltonians. Define $\frac{N_{i}}{N}=a$, $\frac{%
M_{i}}{N}=b$, the energy gap is $\Delta E=\sqrt{1-4b+4ab+4b^{2}-4ab^{2}}$.
In Fig.~$6$, we show $\Delta E$ as a function of $a$ and $b$. By solving the
equation $\frac{\partial \left( \Delta E\right) }{\partial b}=0$, we have $%
b=1/2$. For a given $a$, the minimum of the energy gap $\Delta E$ is at $%
b=1/2$. The minimum of $\Delta E$ is $0$ at $a=0$ and $b=1/2$. In Fig.~$7$,
we show the energy gap as a function of $b$ for $a=0,0.01,0.05$,
respectively. In the algorithm we have to set $M_{i}$ appropriately such
that the point $\left( a,b\right) $ should be far away from the neighborhood
of the point $\left( 0,1/2\right) $.
\begin{figure}[tb]
\centering
\includegraphics[width=0.8\columnwidth, clip]{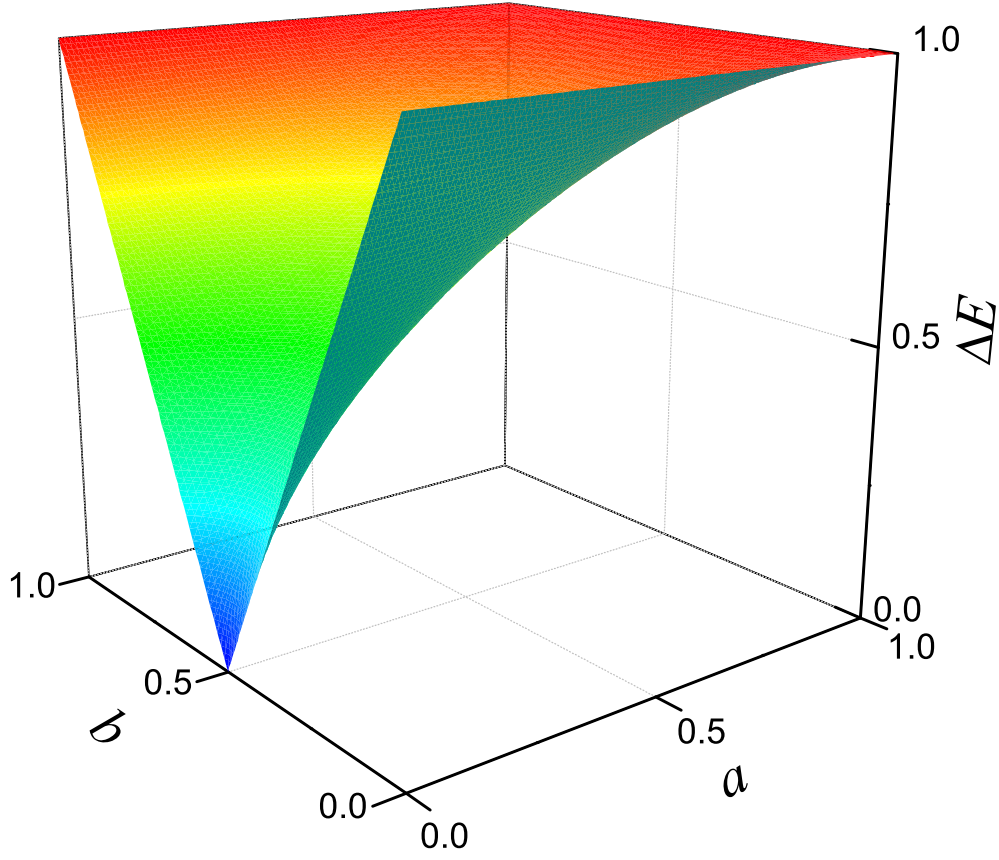}
\caption{ (Color online)~Energy gap between the ground and the first excited
states of the Hamiltonian $H_{i}$ as a function of $a$ and $b$. }
\label{Fig:gap1}
\end{figure}

\begin{figure}[tb]
\centering
\includegraphics[width=0.8\columnwidth, clip]{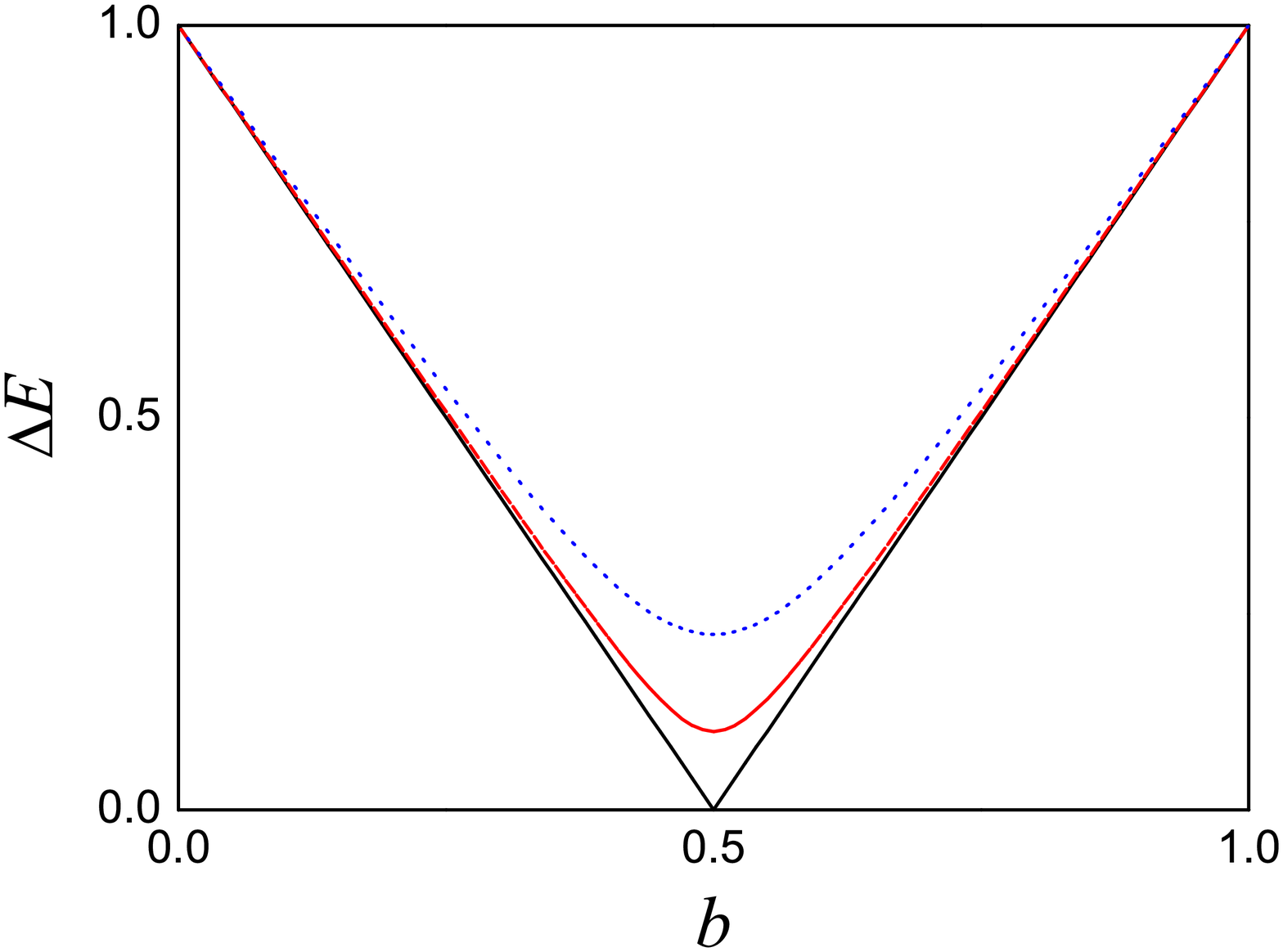}
\caption{ (Color online)~Energy gap between the ground and the first excited
states of the Hamiltonian $H_{i}$ as a function of $b$ by setting $a=0,
0.01, 0.05$, respectively. }
\label{Fig:gap2}
\end{figure}

In our algorithm, $M_{i}$ is an approximate estimation of $N_{i}$ by using
Monte Carlo sampling. Let $b=a+\delta $, where $\delta $ is a small number,
the energy gap can be expanded as
\begin{eqnarray}
\Delta E &=&\sqrt{1-4a+8a^{2}-4a^{3}}-\frac{2\left( 1-a\right) \left(
1-2a\right) }{\sqrt{1-4a+8a^{2}-4a^{3}}}\delta  \notag \\
&&+\frac{2a\left( 1-a\right) }{\left( 1-4a+8a^{2}-4a^{3}\right) ^{3/2}}%
\delta ^{2}+O\left( \delta ^{3}\right) .
\end{eqnarray}%
The first term has a minimum of $\sqrt{11}/3\sqrt{3}\approx 0.638$ at $a=1/3$%
.
\begin{figure}[tb]
\centering
\includegraphics[width=0.83\columnwidth, clip]{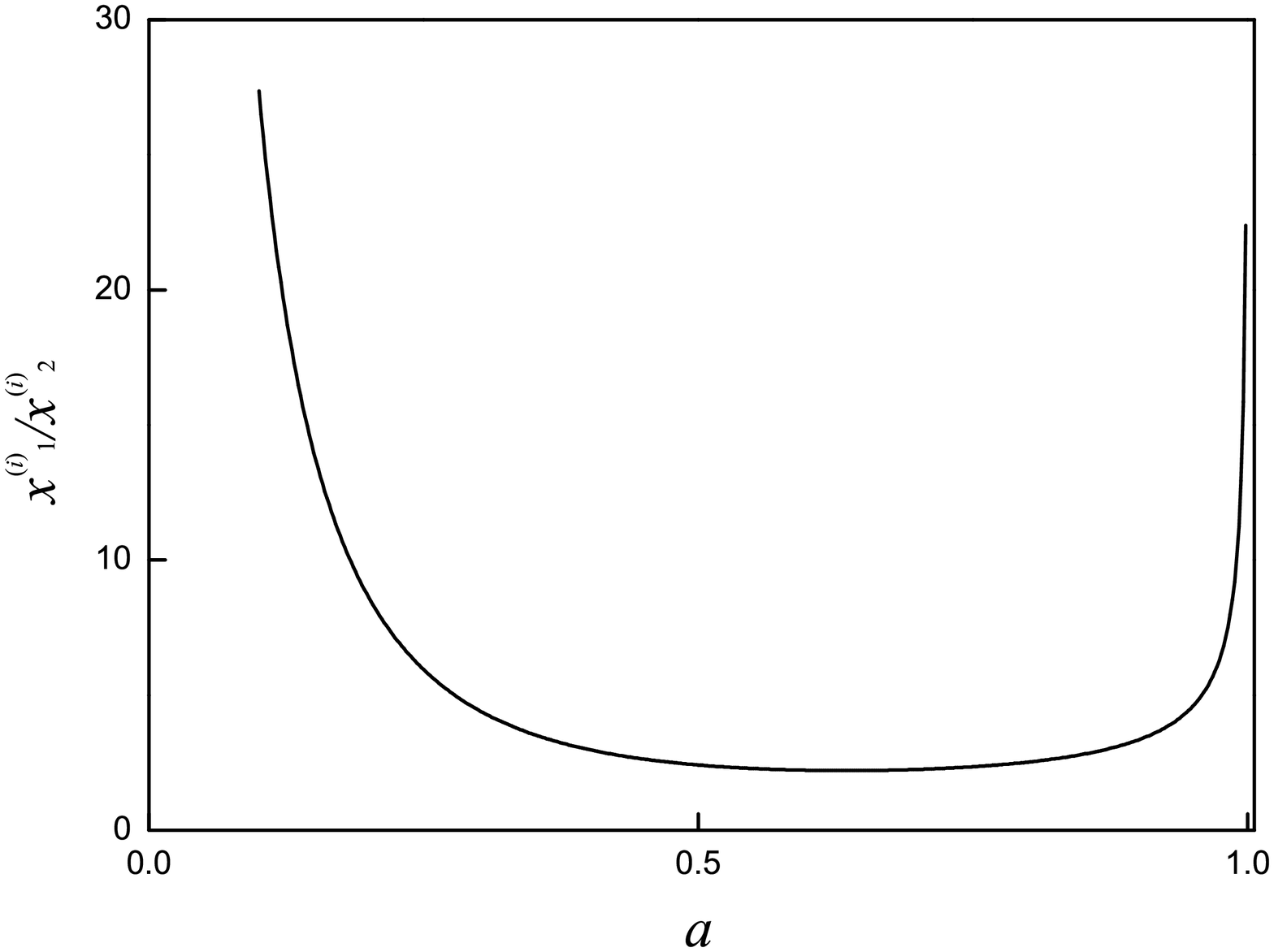}
\caption{Ratio $x_{1}^{(i)}/x_{2}^{(i)}$ between components $x_{1}^{(i)}$
and $x_{2}^{(i)}$ of the ground state of the Hamiltonian $H_{i}$. }
\label{Fig:ratio}
\end{figure}

($ii$) Evaluation of the overlap between the ground states of two adjacent
Hamiltonians. Let $\mathbf{e}=\left( 1,\cdots ,1\right) ^{\text{T}}$ and $%
\mathbf{0}=\left( 0,\cdots ,0\right) ^{\text{T}}$ be $N_{i}\times 1$
vectors, respectively. The ground state of $H_{i}$ is $|V_{-}^{\left(
i\right) }\rangle =x_{1}^{\left( i\right) }\left( \mathbf{e},0\right) ^{%
\text{T}}+x_{2}^{\left( i\right) }\left( \mathbf{0},1\right) ^{\text{T}}$,
where $\left[ x_{1}^{\left( i\right) }\right] ^{2}+\left[ x_{2}^{\left(
i\right) }\right] ^{2}=1$. The components $x_{1}^{\left( i\right) }$ and $%
x_{2}^{\left( i\right) }$ are in the following form:%
\begin{equation}
x_{1}^{(i)}=\frac{1}{A}\left[ \frac{1+\sqrt{\!1-4b-4ab^{2}+4b(a+b)}}{2b\sqrt{%
a\left( 1-a\right) }}-\sqrt{\frac{1}{a}-1}\right].
\end{equation}%
and $x_{2}^{(i)}\!=\frac{1}{A}$, where $A=\sqrt{1+\left[ \frac{1+\sqrt{%
1-4b-4ab^{2}+4b(a+b)}}{2b\sqrt{\!a\left( 1-a\right) }}-\sqrt{\frac{1}{a}-1}%
\right] ^{2}}$.

In basis $\left( \left\{ |q_{i}\rangle \right\} _{q_{i}\in A_{i}}\text{, }%
|q_{i}^{\bot }\rangle \right) $, where $|q_{i}^{\bot }\rangle =\frac{1}{%
\sqrt{N-N_{i}}}\sum_{k\notin A_{i}}|k\rangle $, the state $|V_{-}^{\left(
i\right) }\rangle $ can be written as:
\begin{equation}
|V_{-}^{\left( i\right) }\rangle \!=\!x_{1}^{\left( i\right) }\!\frac{1}{%
\sqrt{N_{i}}}\sum_{q_{i}\in A_{i}}|q_{i}\rangle +x_{2}^{\left( i\right) }\!%
\frac{1}{\sqrt{N-N_{i}}}\sum_{k\notin A_{i}}|k\rangle .
\end{equation}%
Correspondingly, the state $|V_{-}^{\left( i-1\right) }\rangle $ can be
written as:
\begin{widetext}
\begin{eqnarray}
|V_{-}^{\left( i-1\right) }\rangle \! &=&\!x_{1}^{\left( i-1\right) }\!\frac{%
1}{\sqrt{N_{i-1}}}\sum_{q_{i-1}\in A_{i-1}}|q_{i-1}\rangle +x_{2}^{\left(
i-1\right) }\!\frac{1}{\sqrt{N-N_{i-1}}}\sum_{k\notin A_{i-1}}|k\rangle
\notag \\
&=&\!x_{1}^{\left( i-1\right) }\!\frac{1}{\sqrt{N_{i-1}}}\left( \sum_{k\in
A_{i-1}\setminus A_{i}}|k\rangle +\sum_{k\in A_{i}}|k\rangle \right)
+x_{2}^{\left( i-1\right) }\!\frac{1}{\sqrt{N-N_{i-1}}}\sum_{k\notin
A_{i-1}}|k\rangle.
\end{eqnarray}
\end{widetext}
Thus the overlap between the ground states $|V_{-}^{\left( i-1\right)
}\rangle $ and $|V_{-}^{\left( i\right) }\rangle $ is
\begin{widetext}
\begin{eqnarray}
g_{0}^{(i)}\! &=&\!\langle V_{-}^{\left( i\!-\!1\right) }|V_{-}^{\left(
i\right) }\rangle \!=\!\sqrt{\!\frac{N_{i}}{N_{i-1}}}x_{1}^{\left(
i\!-\!1\right) \ast }x_{1}^{\left( i\right) }\!+ \!\frac{N_{i\!-\!1}\!-\!N_{i}}{\sqrt{\!N_{i\!-\!1}\!(N\!-\!N_{i})}}%
x_{1}^{\left( i\!-\!1\right) \ast }x_{2}^{(i)}\!+\!\sqrt{\frac{%
N\!-\!N_{i\!-\!1}}{N\!-\!N_{i}}}x_{2}^{\left( i\!-\!1\right) \ast
}x_{2}^{\left( i\right) }.
\end{eqnarray}
\end{widetext}

By setting $b=a+\delta $, the ratio $x_{1}^{(i)}/x_{2}^{(i)}$ can be
expanded as%
\begin{widetext}
\begin{eqnarray}
x_{1}^{(i)}/x_{2}^{(i)} &=&\frac{-2a+2a^{2}+1+\sqrt{\!1-4a+8a^{2}-4a^{3}}}{2a%
\sqrt{a\left( 1-a\right) }}+ \frac{-1\!+\!2a\!-\!2a^{2}\!-\!\sqrt{1\!-\!4a\!+\!8a^{2}\!-\!4a^{3}}}{%
2a^{2}\!\sqrt{a\left( 1\!-\!a\right) }\!\sqrt{\!1-\!4a\!+\!8a^{2}\!-\!4a^{3}}}%
\delta \!+\! O\left( \delta^{2} \right) .
\end{eqnarray}
\end{widetext}
The first term of the above expansion is shown in Fig.~$8$, which has a
minimum of $2.21$ at $a=\left( 3-\sqrt{3}\right) /2$.

In order to make the overlap $g_{0}^{(i)}$ not to be exponentially small, we
require that $x_{1}^{\left( i\right) }>x_{2}^{\left( i\right) }$, such that
the overlap $g_{0}^{(i)}$ is guaranteed to be polynomial large when the
ratio $N_{i-1}/N_{i}$ is polynomial large. This can be achieved by setting $%
M_{i}$ or $b$ appropriately. By solving the inequality $x_{1}^{\left(
i\right) }>x_{2}^{\left( i\right) }$, we have $b<\frac{1}{2\left( 1-a\right)
}$, which also can be written as $2M_{i}\left( N-N_{i}\right) <N^{2}$. Both $%
M_{i}$ and $N_{i}$ are in decreasing order with the increasing of the steps
of the algorithm, therefore the condition is easily satisfied in the last
few steps of the algorithm. While at the beginning steps, $N_{i}$ can be
obtained approximately by using the Monte Carlo sampling. Then we can set $%
M_{i}$ accordingly to satisfy the condition.

Summarizing the above calculation results, we find that the parameters $%
M_{i} $ should be set such that: the point $\left( a,b\right) $ should be
far away from the neighborhood of the point $\left( 0,1/2\right) $, and $%
2M_{i}\left( N-N_{i}\right) <N^{2}$.

\end{appendix}

\end{document}